# THERMODYNAMICS OF MIXTURES CONTAINING A STRONGLY POLAR COMPOUND. 9. LIQUID-LIQUID EQUILIBRIA FOR $\varepsilon$-CAPROLACTAM + SELECTED ALKANES

VÍCTOR ALONSO; IVÁN ALONSO, ISMAEL MOZO, JUAN ANTONIO GONZÁLEZ*, ISAÍAS GARCÍA DE LA FUENTE AND JOSÉ CARLOS COBOS

**G.E.T.E.F., Grupo Especializado en Termodinámica de Equilibrio entre Fases,** Departamento de Física Aplicada, Facultad de Ciencias, Universidad de Valladolid, E-47071, Valladolid (Spain).

*corresponding author
E-mail: jagl@termo.uva.es

**ABSTRACT**

The coexistence curves of the liquid-liquid equilibria (LLE) for systems of $\varepsilon$-caprolactam with heptane, octane, nonane, decane or 2,2,4-trimethylpentane have been determined by the method of the critical opalescence using a laser scattering technique. All the curves show an upper critical solution temperature (UCST), have a rather horizontal top and their symmetry depends on the size of the alkane. The UCST increases almost linearly with the chain length of the $n$-alkane. For the octane mixture, the UCST is lower than for the solution including 2,2,4-trimethylpentane.

1. **INTRODUCTION**

We are engaged in a systematic investigation on the thermodynamic properties of mixtures containing a compound with a very high dipolar moment in gas phase ($\mu$), such as, thiophene 1,1-dioxide[1,2] ($\mu$ = 16.04 $10^{-30}$ C·m [3]); dimethyl sulfoxide[4] ($\mu$ = 13.54 $10^{-30}$ C·m [3]); 1-methyl-2-pyrrolidinone[5,6] (NMP; $\mu$ = 13.64 $10^{-30}$ C·m [3]); *N,N*-dialkylamides[6,7] (*N,N*-dimethylmethanamide (DMF), $\mu$ = 12.28 $10^{-30}$ C·m[8]; *N,N*-dimethyltacetamide (DMA), $\mu$ = 12.37 $10^{-30}$ C·m [8]); *N*-alkylamides[9] or propylene carbonate ($\mu$ = 16.49 $10^{-30}$ C·m [3]).

Amides, amino acids, peptides and their derivatives are of interest because they are simple models in biochemistry. N-methylmethanamide possesses the basic (−CO) and acidic (−NH) groups of the very common, in nature, peptide bond[10]. So, proteins are polymers of amino acids linked to each other by peptide bonds. Cyclic amides are also of importance due to they are related to structural problems in biochemistry. Consequently, the understanding of liquid mixtures involving the amide functional group is necessary as a first step to a better knowledge of complex molecules of biological interest[11]. For example, the aqueous solution of DMF is a model solvent representing the environment of the interior of proteins. On the other hand, DMF and NMP are used as highly selective extractants for the recovery of aromatic and saturated hydrocarbons from petroleum feedstocks[12], and $\varepsilon$-caprolactam is used for the production of nylon 6, which is a polycaprolactam formed by a ring-opening polymerization.

From a theoretical point of view, amides are also a very interesting class of compounds. In pure liquid state, they present a significant local order[13] as their quite high heats of vaporization indicate[14]. In the case of *N,N*-dialkylamides, this is due to the dominance of the general dipole-dipole interactions[13], which can be ascribed to their very high effective dipole moments[7], a useful magnitude to examine the impact of polarity on bulk properties[7,15] For primary and secondary amides, their self-association via H-bonds must be also taken into account[13,16]. This is the case of $\varepsilon$-caprolactam, a cyclic secondary amide with a high dipole moment (12.97 $10^{-30}$ C·m[8]).

As a continuation of our investigations on mixtures involving amides[5-7,9,17], we report here LLE curves for $\varepsilon$-caprolactam with heptane, octane, nonane, decane or 2,2,4-trimethylpentane. As far as we know, no LLE data for these systems are available in the literature.

## 2. EXPERIMENTAL

*2.1 Materials*

Heptane (142-82-5, puriss p.a. ≥ 0.995); octane (11-65-9, puriss p.a. ≥ 0.99); nonane (111-84-2, puriss p.a. 0.99); decane (124-18-5, puriss p.a. ≥ 0.99) and 2,2,4-trimethylpentane (540-84-1, puriss p.a. ≥ 0.995) were from Fluka (purities expressed in mass fraction). $\varepsilon$-Caprolactam (106-60-2, puriss p.a. ≥ 0.99) was from Aldrich. Prior to the measurements, the chemicals were stored over molecular sieves (Union Carbide Type 4Å from Fluka). All these chemicals were used without other further treatment. The densities ρ at 298.15 K and atmospheric pressure were in good agreement with literature values (Table 1). The fusion point of $\varepsilon$-Caprolactam (342.21 K) measured in our laboratory is in excellent agreement with the value reported in the literature (342.3 K).[18] The water contents, determined by the Karl-Fischer method, were as follows (in mass fraction): $2 \times 10^{-5}$ for heptane, octane and $\varepsilon$-caprolactam, $10^{-5}$ for nonane and decane and $5 \times 10^{-5}$ for 2,2,4-trimethylpentane.

*2.2 Apparatus and Procedure*

Mixtures were prepared by mass, with a weighing accuracy to ± 0.00001 g, in Pyrex tubes of 0.9 cm i.d. and about 4 cm length, which then were immediately sealed by capping at atmospheric pressure and room temperature. Conversion to molar quantities was based on the relative atomic mass table of 1985 issued by IUPAC in 1986[19].

The coexistence curves of the binary mixtures were determined by the method of the critical opalescence. The samples in the sealed Pyrex tubes are placed in a thermostat bath a few hundredths of degree above the expected temperature. A He-Ne laser is placed on one side of the equilibrium cell so that the light beam passing through the solution is focused on a light sensor. When decreasing slowly the temperature (1.2 K h$^{-1}$), the growth of small drops of the dispersed liquid phase causes a diffusion of light during the transition. This results in a voltage variation, measured by a digital multimeter model Agilent 34410A connected to a PC, which allows simultaneous accurate measurements of the transition temperatures. The temperature was measured with a precision of ± 0.001 K and estimated accuracy of ± 0.05 K by a Pt-1000 resistance connected to a multimeter model Philips PM2353, in such way that the resistance variations upon cooling are also registered in a PC. The thermometer was calibrated on the basis of the ITS-90 scale of temperature using the triple point of the water, and the fusion point of Ga. The separation temperatures were reproducible to ± 0.02 K for temperatures near the upper critical solution temperature. The precision of the equilibrium composition is expected to be better than 0.0005 in mole fraction. The weighing technique gives a precision better than 0.0001 in mole fraction, but this is reduced slightly due to partial evaporation of the more volatile component to the free volume of the ampoule (≈1.17 cm$^3$).

## 3. RESULTS

Table 2 lists the direct experimental results of the liquid-liquid equilibrium temperatures $T$ vs. the mole fraction of the amide, $x_1$, for the investigated mixtures (Fig. 1,).

All the systems show an UCST. The LLE coexistence curves have a flat maximum, and their symmetry depends on the size of the alkane (Fig. 1).

The coordinates of the critical points, $x_{1c}$ and $T_c$ (Table 3), were obtained by reducing the experimental data with the equation[20,21]

$$T/K = T_c/K + k|y - y_c|^m \tag{1}$$

where

$$y = \frac{\alpha\, x_1}{1 + x_1(\alpha - 1)} \tag{2}$$

$$y_c = \frac{\alpha\, x_{1c}}{1 + x_{1c}(\alpha - 1)} \tag{3}$$

In eqs 1 to 3, $m$, $k$, $\alpha$, $T_c$ and $x_{1c}$ are the coefficients to be fitted to the experimental results. When $\alpha = 1$, eq 1 is similar to the well-known equation:[15,22,23]

$$\Delta \lambda = B \tau^\beta \tag{4}$$

where $\Delta \lambda_1 = \lambda_1' - \lambda_2''$ is the so-called order parameter, which can be any density variable in the conjugate phase (in our case $\lambda_1 = x_1$), $\tau$ is the reduced temperature $(T_c - T)/T_c$ and $\beta$ a critical exponent corresponding to this order parameter. The $\beta$ value depends on the theory applied to its determination.[22,23] More details are given elsewhere.[24]

The fitting was performed using the Marquardt algorithm[25] with all the points weighted equally. Results are collected in Table 3. Also listed is the standard deviation defined by:

$$(\sigma(T)/K) = \left[\sum \left(T_i^{\exp} - T_i^{cal}\right)^2 /(N-n)\right]^{1/2} \tag{5}$$

where $N$ and $n$ stand for the number of data points and the number of fitted parameters, respectively. We note that eq 1 fits well the experimental data.

## 4. DISCUSSION

The UCST of the studied systems increases with the chain length of the *n*-alkane (Fig. 2). The same behaviour is observed in mixtures formed by *n*-alkane with, DMF[17], DMA[26-30], NMP[5] (Fig. 3), linear alkanone[31], linear organic carbonate[32], acetic anhydride[33], alkoxyethanol[24,34,35] or polyether[36,37]. Fig. 1 shows that the LLE curves are progressively skewed to high $x_1$ values when the chain length of the alkane increases. A similar trend is encountered for many other mixtures.[24,26-37] On the other hand, we note that UCST(2,2,4-trimethylpentane) > UCST(octane) (Table 3, Fig.1). UCST values of systems with the same solute and different isomeric alkanes strongly depend on the solute nature and size and shape of the alkanes. So, when methanol is mixed with isomeric hexanes, we have the following UCST values:[38] 313.42 K (hexane); 304.05 K (2-methylpentane); 293.02 K (2,2-dimethylbutane); 299.20 (2,3-dimethylbutane). In the case of 2-methoxyethanol solutions,[39] the UCST is 327.94 K and 322.41 K for the mixtures including octane, and 2,2,4-trimethylpentane, respectively. In systems with polystyrene1241,[40] the UCST is 285.2 K for the octane mixture, and 262.2 K for that including 2,2,4-trimethylpentane. For the same solute and isomeric hexanes, we have 274.9 K (hexane); 278.2 K (2,2,dimethylbutane); 310.4 K (2,3-dimethylbutane. The UCST values are 322.41 K, 327.60 K and 319.22 K when NMP is mixed with hexane[41], 2,2-dimethylbutane[42], or 2,3-dimethylbutane[42], respectively.

Finally, it is interesting to note that interactions between amide molecules are stronger in $\varepsilon$-caprolactam systems than in those mixtures including tertiary *N,N*-dialkylamides (DMF, DMA, NMP), as it is indicated by the higher UCSTs of the former solutions (Fig.2 ). This may be due to the existence of H-bonds between $\varepsilon$-caprolactam molecules.

## 5. CONCLUSIONS

LLE coexistence curves were determined for mixtures of $\varepsilon$-caprolactam with hexane, heptane, octane, nonane, decane or 2,2,4-trimethylpentane. The UCST increases almost linearly with the chain length of the *n*-alkane, and is higher for the solution including 2,2,4-trimethylpentane than for the octane system.

**ACKNOWLEDGEMENTS**

The authors gratefully acknowledge the financial support received from the Consejería de Educación y Cultura of Junta de Castilla y León, under Projects VA075A07 and VA052A09 and from the Ministerio de Educación y Ciencia, under the Project FIS2007-61833. V.A. and I.A. also gratefully acknowledge the grants received from the Universidad de Valladolid, and the Junta de Castilla y León, respectively.

TABLE 1

Experimental liquid-liquid equilibrium temperatures for $\varepsilon$-caprolactam(1) + alkane(2) mixtures.

| $x_1$ | $T$/K | $x_1$ | $T$/K |
|---|---|---|---|
| \multicolumn{4}{c}{$\varepsilon$-caprolactam(1) + heptane(2)} | | | |
| 0.2691 | 346.47 | 0.4991 | 352.17 |
| 0.2952 | 348.31 | 0.5217 | 352.08 |
| 0.3188 | 349.57 | 0.5464 | 351.92 |
| 0.3459 | 350.69 | 0.5709 | 351.62 |
| 0.3690 | 351.29 | 0.5962 | 351.10 |
| 0.3968 | 351.72 | 0.6204 | 350.34 |
| 0.4207 | 352.00 | 0.6310 | 349.83 |
| 0.4452 | 352.14 | 0.6488 | 348.84 |
| 0.4700 | 352.21 | 0.6702 | 347.44 |
| 0.4977 | 352.18 | 0.6955 | 344.80 |
| 0.4978 | 352.11 | | |
| \multicolumn{4}{c}{$\varepsilon$-caprolactam (1) + octane(2)} | | | |
| 0.3058 | 349.77 | 0.4994 | 354.57 |
| 0.3260 | 350.88 | 0.5240 | 354.63 |
| 0.3479 | 351.98 | 0.5377 | 354.58 |
| 0.3614 | 352.57 | 0.5552 | 354.55 |
| 0.3777 | 353.17 | 0.5751 | 354.48 |
| 0.4007 | 353.72 | 0.5986 | 354.29 |
| 0.4139 | 354.02 | 0.6303 | 353.69 |
| 0.4264 | 354.30 | 0.6503 | 353.18 |
| 0.4457 | 354.36 | 0.6666 | 352.48 |
| 0.4618 | 354.55 | 0.6766 | 352.06 |
| 0.4772 | 354.58 | 0.6930 | 351.01 |
| 0.4788 | 354.56 | | |
| \multicolumn{4}{c}{$\varepsilon$-caprolactam(1) + nonane(2)} | | | |
| 0.3442 | 354.21 | 0.5715 | 358.60 |
| 0.3732 | 355.78 | 0.5865 | 358.56 |
| 0.3957 | 356.75 | 0.5963 | 358.57 |
| 0.4188 | 357.45 | 0.6221 | 358.45 |
| 0.4477 | 358.01 | 0.6455 | 358.14 |
| 0.4699 | 358.31 | 0.6551 | 357.99 |

TABLE 2 (continued)

| | | | |
|---|---|---|---|
| 0.5007 | 358.49 | 0.6725 | 357.62 |
| 0.5059 | 358.53 | 0.6820 | 357.30 |
| 0.5209 | 358.60 | 0.7062 | 356.20 |
| 0.5449 | 358.62 | 0.7207 | 355.41 |
| 0.5488 | 358.70 | | |
| $\varepsilon$-caprolactam(1) + decane(2) | | | |
| 0.3936 | 360.15 | 0.5565 | 363.41 |
| 0.4175 | 361.23 | 0.5756 | 363.43 |
| 0.4430 | 362.01 | 0.5957 | 363.45 |
| 0.4707 | 362.65 | 0.6042 | 363.45 |
| 0.4769 | 362.78 | 0.6160 | 363.39 |
| 0.4945 | 363.09 | 0.6459 | 363.32 |
| 0.4969 | 363.06 | 0.6711 | 363.06 |
| 0.4980 | 363.09 | 0.6971 | 362.59 |
| 0.5159 | 363.24 | 0.7207 | 361.65 |
| 0.5449 | 363.40 | 0.7421 | 360.76 |
| $\varepsilon$-caprolactam(1) + 2,2,4-trimethylpentane(2) | | | |
| 0.3206 | 358.94 | 0.4999 | 362.34 |
| 0.3476 | 360.11 | 0.5052 | 362.35 |
| 0.3681 | 360.84 | 0.5213 | 362.35 |
| 0.3947 | 361.59 | 0.5315 | 362.31 |
| 0.3971 | 361.60 | 0.5460 | 362.27 |
| 0.4198 | 361.94 | 0.5765 | 362.06 |
| 0.4486 | 362.19 | 0.5951 | 361.80 |
| 0.4733 | 362.30 | 0.6222 | 361.21 |
| 0.4766 | 362.34 | 0.6472 | 360.26 |
| 0.4865 | 362.39 | 0.6717 | 358.93 |
| 0.4879 | 362.34 | | |

TABLE 2

Coefficients in eq. (1) for the fitting of the ($x_1$, $T$) pairs given in Table 2 for $\varepsilon$-caprolactam (1) + alkane(2) mixtures; $\sigma$ is the standard deviation defined by eq 5.

| $N^a$ | $m$ | $K$ | $\alpha$ | $T_c$/K | $x_{1c}$ | $\sigma$/K |
|---|---|---|---|---|---|---|
| $\varepsilon$-caprolactam (1) + heptane(2) | | | | | | |
| 21 | 3.10 | −811 | 0.830 | 352.13 | 0.482 | 0.06 |
| $\varepsilon$-caprolactam(1) + octane(2) | | | | | | |
| 23 | 3.05 | −653 | 0.781 | 354.51 | 0.519 | 0.05 |
| $\varepsilon$-caprolactam(1) + nonane(2) | | | | | | |
| 21 | 3.18 | −774 | 0.696 | 358.61 | 0.554 | 0.04 |
| $\varepsilon$-caprolactam(1) + decane(2) | | | | | | |
| 20 | 3.03 | −560 | 0.657 | 363.43 | 0.586 | 0.04 |
| $\varepsilon$-caprolactam(1) + 2,2,4-trimethylpentane(2) | | | | | | |
| 21 | 2.93 | −593 | 0.746 | 362.34 | 0.505 | 0.03 |

[a] number of experimental data points.

**CAPTION TO FIGURES**

**FIG. 1** LLE of $\varepsilon$-caprolactam(1) + alkanes (2) mixtures. Points, experimental results (this work): (●), heptane; (■), octane; (▲), nonane; (▼), decane; (◆), 2,2,4-trimethylpentane. Solid lines, results from the fitting equation (1).

**FIG. 2** Upper critical solution temperatures $T_c$, vs. $n$, the number of carbon atoms in the $n$-alkane, for some amide + $n$-alkane mixtures. Values for systems with DMA were taken from references 26-30. For NMP mixtures, see reference 5, and for DMF solutions, references 17,43,44.

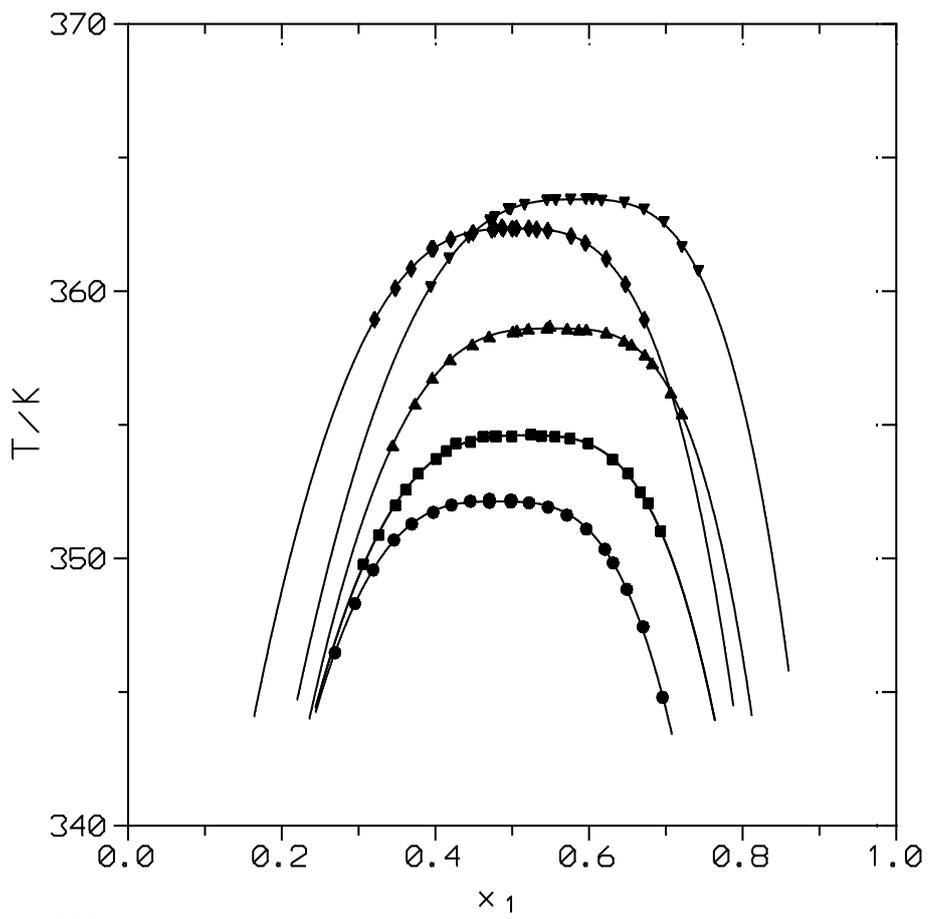

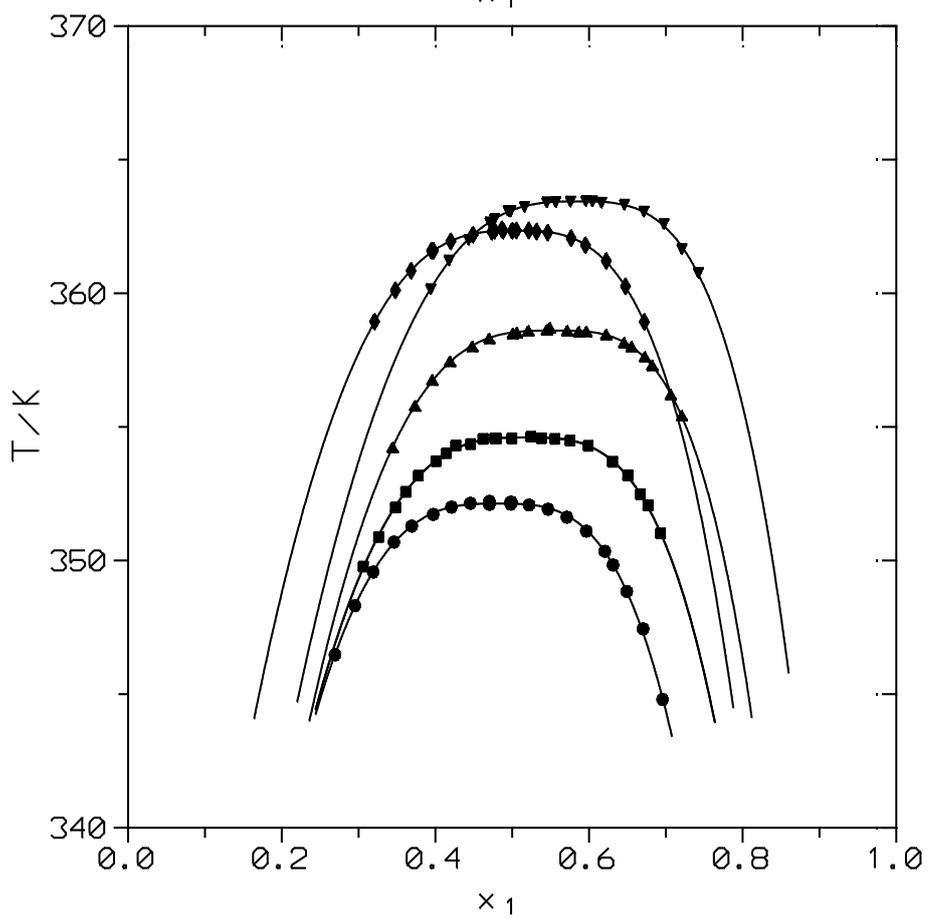

FIG. 1

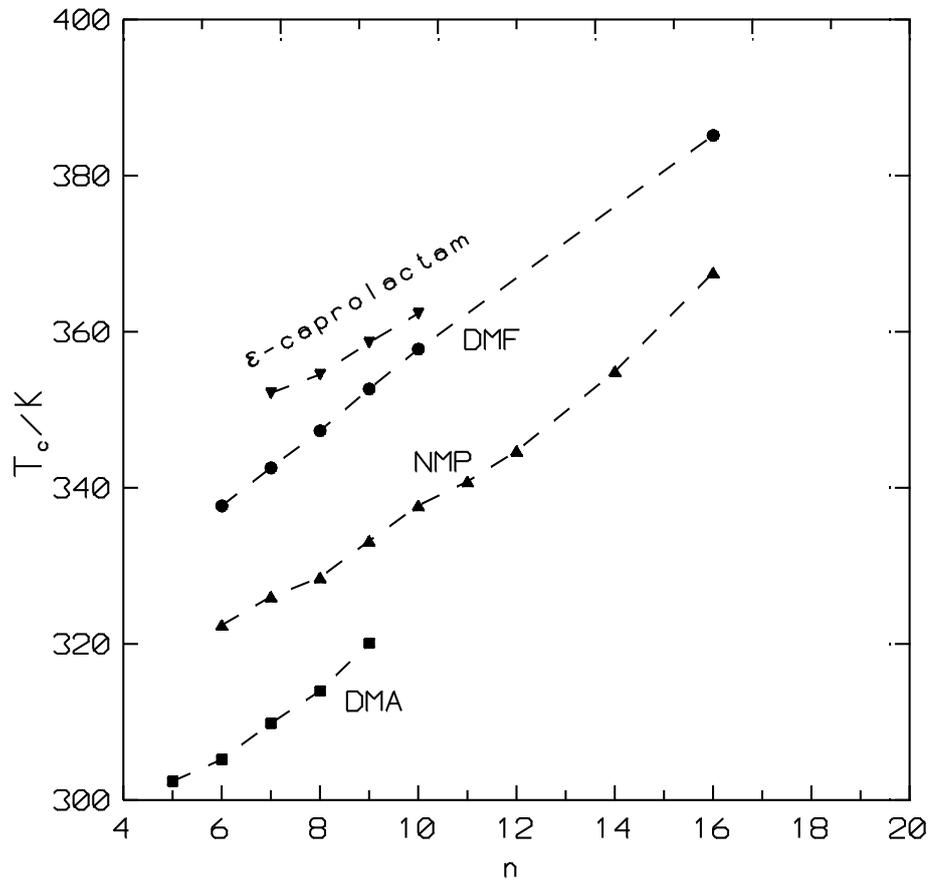

FIG. 2